\documentclass[12pt]{JHEP3}

\usepackage{enumerate}
\usepackage{amsbsy}
\usepackage[dvips]{graphicx}
\usepackage{graphics}

\newcommand{\be}{\begin{equation}} \newcommand{\ee}{\end{equation}}
\newcommand{\bea}{\begin{eqnarray}} \newcommand{\eea}{\end{eqnarray}}

\newcommand{\para}{\paragraph}

\renewcommand{\a}{\alpha}

\newcommand{\nonum}{\\}

\newcommand{\PRD}[1]{{\it Phys. Rev.} {\bf D#1}}

\newcommand{\APJ}[1]{{\it Astrophys. J.} {\bf #1}}

\newcommand{\GRG}[1]{{\it Gen. Rel. Grav.} {\bf #1}}

\newcommand{\IJMPD}[1]{{\it Int. J. Mod. Phys.} {\bf D#1}}

\title{Vacuum energy and dynamical symmetry breaking in curved spacetime}

\author{Syksy R\"{a}s\"{a}nen\thanks{Essay written for the Gravity Research Foundation 2012 Awards for Essays on Gravitation. Submitted on March 28.} \\

University of Helsinki, Department of Physics \\
and Helsinki Institute of Physics \\
P.O. Box 64, FIN-00014 University of Helsinki, Finland \\

\email{syksy {\it dot} rasanen {\it at} iki {\it dot} fi}}

\abstract{
We argue that calculating vacuum energy requires
quantum field theory whose axioms are adapted to
curved spacetime.
In this context, we suggest that non-zero
vacuum energy is connected to dynamical breaking
of electroweak symmetry.
The observed meV scale can be understood in terms
of electroweak physics via a naive estimate.
The scenario requires all particle masses to have a
dynamical origin. Any Higgs particle
has to be a composite, and the origin of vacuum
energy might be probed at the LHC.
}

\preprint{HIP-2012-08/TH}

\begin{document}
  
\setcounter{tocdepth}{2}

\setcounter{secnumdepth}{3}



\para{The vacuum energy problem.}

Vacuum energy has a long history \cite{Peebles:2002, Kragh:2011},
and it has been brought to focus by cosmological observations which
have in the past two decades established that the universe expands
faster than expected.
The usual view of the vacuum energy problem\footnote{The terms vacuum
energy and cosmological constant are often used interchangeably.
The cosmological constant is a geometrical term in the
classical Einstein equation, and it changes the way
spacetime curves in the absence of matter.
Vacuum energy has its origin in quantum theory, and it
is a form of matter which has the same effect as the
cosmological constant. We only discuss vacuum energy.}
is that quantum field theory (QFT) predicts vacuum energy,
cosmological observations reveal its presence, and there
is a huge discrepancy between the predicted and
measured values.

One measure of the perceived severity of the problem
is the popularity of anthropic arguments.
When no solution that would determine the value of
a physical quantity has been found, it may be tempting
to take this as evidence that there is no solution, and that the
measured value is due to environmental contingency.
This road was earlier taken with regard to spatial curvature
\cite{Collins:1973}, though anthropic arguments were eventually left
by the wayside when a dynamical, testable, explanation was found
with the introduction of inflation.
The appeal to anthropic arguments indicates that the problem
does not involve contradiction between theory and observation,
nor theoretical inconsistency, simply unmet expectations.
In the vacuum energy case, it is useful to consider these
expectations carefully, because the simple summary given
above is somewhat misleading.

\para{No vacuum energy in flat spacetime QFT.}

First of all, the interpretation
of the observations in terms of vacuum energy is not beyond
reasonable doubt; apart from explanations involving
exotic matter or modifications to gravity,
it is possible that the change in the expansion rate
is simply due to the known breakdown of homogeneity and isotropy
at late times \cite{Rasanen:2011, Buchert:2011}. 
Nevertheless, even if the cosmological observations were explained
by something else than vacuum energy, the question of why vacuum
energy is not large would remain. Let us assume that the correct
explanation for the observations is indeed vacuum energy, and look at
the theoretical side.

It is often claimed that the Casimir effect shows that
vacuum energy is real \cite{Peebles:2002}.
However, as discussed in \cite{Jaffe:2005}, the Casimir
effect can be calculated without any reference to vacuum
energy, and gives no more indication of the reality of vacuum
energy than any other loop contribution in QFT. Indeed,
the Lamb shift was quoted to the same effect in the
background material for the 2011 Nobel prize in physics, awarded
for the discovery of accelerating expansion \cite{Nobel}.
However, while loop effects may be related to vacuum
fluctuations, these are distinct from vacuum energy\footnote{In
particular, referring to vacuum energy as the energy of fluctuations
of the vacuum is misleading, because the vacuum is an energy eigenstate,
so there are no fluctuations in the energy.}.

QFT formulated in flat spacetime is sensitive only to
differences in energy: an arbitrary constant can be added to the
Lagrangian without changing the physics, so QFTs
which differ only by a constant energy are equivalent.
As the concept of absolute energy is not part of the physical
content of flat spacetime QFT, it is even in principle impossible
to obtain any prediction about the value of vacuum energy, any more
than it is possible to predict gauge-fixing parameters.

Arguments about vacuum energy in flat spacetime
are therefore at most suggestive, and suggestions are easy
to misconstrue. For example, it has been argued that calculating
vacuum energy by integrating over momentum space modes with a
cut-off $\Lambda$ much larger than the masses present in the theory
shows that it is of the order $\Lambda^4$.
However, the resulting energy-momentum tensor is traceless, so it
behaves like radiation and does not lead to accelerating
expansion, as is well known \cite{Peebles:2002, Hollands:2004}.
When the integration range is extended to infinity, the equation
of state becomes indefinite as the result diverges.

In any case, vacuum expectation values in QFT do not reduce to simple
momentum mode decomposition \cite{Hollands:2004}. From the field theory
point of view, vacuum energy is a renormalisable term (supersymmetric
theories excepted), and its value is arbitrary, like particle masses.
However, whereas masses are fixed by observation, vacuum energy is not
observable in flat spacetime QFT.
If spacetime is curved, the situation is different, since
gravity responds to absolute and not relative amounts of energy.
The new piece of information which has to be supplied
is how quantum fields couple to gravity.

\para{Symmetry breaking in curved spacetime QFT.}

The effect of quantum fields on spacetime curvature can be
studied in semiclassical quantum gravity, where spacetime is treated
classically and its dynamics is sourced by the expectation value
of the energy-momentum tensor. Vacuum energy
then becomes a physical quantity with arbitrary magnitude.
However, the vacuum state can evolve as the universe expands.
A well-known problem is that changes in the vacuum energy
associated with the Higgs field are of the order of the electroweak
scale. While vacuum energy may be simply set to the value that
would explain the late-time observations, this does not offer any
insight into why it has this value.

In the above discussion it has been assumed that we can use the
rules of flat spacetime QFT to calculate vacuum expectation values.
However, the structure of QFT in flat spacetime is rooted in special
properties of Minkowski space which curved spacetime does not share.
The crux of the vacuum energy issue is the relation between
the energy-momentum tensor and spacetime curvature, so we should
consider how the situation changes when QFT is formulated
consistently in curved spacetime.

In QFT based on axioms suited to curved
spacetime, the vacuum energy of a massive free scalar
field is arbitrary for non-zero mass, but necessarily
vanishes for zero mass, in contrast to the flat
spacetime theory, in which the vacuum energy is always
arbitrary \cite{Hollands:2008a, Hollands:2008b}.
Vacuum expectation values are expected to
depend non-analytically on the parameters of the theory,
and it has been suggested that this could lead to
vacuum energy much smaller than the natural scale of the theory,
and that this might be relevant for present-day cosmic acceleration
\cite{Hollands:2008b}.

To elaborate further, we can propose the following scenario.
Since the Standard Model Higgs field is massive, it might seem
that the above result is not relevant.
However, if there is no elementary Higgs field,
the situation changes \cite{Rasanen:2011}.
If it is also true for fermions and vector fields
that vacuum energy vanishes for zero mass and
there are no elementary scalars, so that all masses are
dynamically generated, then this would explain why the
vacuum energy is zero. (This is sometimes known as the
``old'' vacuum energy problem.) Two questions immediately arise.
What happens when interactions are included, in particular
those which dynamically generate masses and composite particles?
And second, is it possible to understand the value of vacuum
energy which would explain the observations?
These issues turn out to be related.

It may be that as the symmetry which keeps particles at zero mass
is dynamically broken, it can no longer enforce vanishing
vacuum energy. Vacuum energy should thus vanish as coupling
constants go to zero, the theory becomes free and symmetry
is restored. Vacuum energy should also depend non-analytically
on the couplings. For a coupling constant $g$, perhaps the
simplest possibility is $e^{-1/g^2}$.
In electroweak physics, the scale of particle masses is
related to the Higgs vacuum expectation value $v\approx246$ GeV.
Taking $g^2=\a\approx1/137$, the vacuum energy density could
then naively be expected to be
\bea
  \rho_\mathrm{vac} \sim e^{-1/g^2} v^4 \approx (0.33\ \mathrm{meV})^4 \ .
\eea

\noindent This is close to the value that would explain the observations,
$\rho_\mathrm{vac}=3 H^2 M_\mathrm{Pl}^2 \Omega_\Lambda \approx (h/0.7)^2 (\Omega_\Lambda/0.7)\ (2.3$ meV)$^4$.
The estimate is exponentially sensitive to inserting $e^2=4\pi\a$
instead of $\a$ and to the presence of other factors such as weak
mixing angles, and the functional form is mere conjecture.
Nevertheless, this does show that the scale which would explain
late-time cosmological observations may emerge naturally from
electroweak physics in curved spacetime.

We can outline steps to take to see whether the above
conjecture holds any truth.
First we should consider the vacuum energy of
free fermions and vector bosons in curved spacetime QFT,
then introduce a model of symmetry breaking and
trace what happens to vacuum energy.
Eventually, a full model of dynamical mass generation
should be considered: realistic models tend to have a new scale
in addition to the electroweak scale, and the dynamics
can be rather involved \cite{composite}.

It may not be necessary to deal with spacetime curvature.
The key ingredient is the new set of axioms:
this is not flat spacetime QFT, but the flat spacetime limit
of curved spacetime QFT. The experimental
success of QFT based on flat spacetime axioms shows that the curved
spacetime theory must reduce extremely accurately to the flat
spacetime version as far as non-gravitational physics is concerned.
However, when it comes to effects which are absent or ambiguous
in the flat spacetime theory and to which particle physics experiments
are not sensitive, the results can be very different.

\para{Conclusions.}

If vacuum energy is the cause of the increased expansion rate
at late times, its value is tiny compared to particle
physics scales. A natural explanation would be that there is
a symmetry which sets vacuum energy exactly to zero, and which
is slightly broken. In the present scenario, the symmetry is related
to particles having zero mass, and it is broken by the emergence of
new effective degrees of freedom in dynamical breaking of electroweak
symmetry. Vacuum energy is thus, like massive particles, an
emergent phenomenon.
Dynamical mass generation is not a small effect, but its impact on
vacuum energy is diluted exponentially by the non-analytical
dependence of vacuum expectation values on coupling constants.

The scenario requires that there are no massive elementary
scalar fields, in particular that any Higgs particle is a
composite, which is a qualitative prediction for collider physics.
As vacuum energy is related to the spectrum of masses of
particles at the electroweak scale, the vacuum energy problem
is connected to the Higgs mass hierarchy problem and
it may be possible to probe it at the LHC.
It is also necessary that there are no particles with masses
larger than the electroweak scale, or that the contribution
of such particles to vacuum energy is correspondingly more
suppressed. On the theoretical side,
the dynamical generation of a mass scale and the choice of the
axioms of QFT relates vacuum energy to the Millennium problem
of defining Yang-Mills theory and understanding the mass gap \cite{Millennium}.

Even if the idea turns out to be wrong, it shows that
addressing the vacuum energy problem does not necessarily require
delving into speculative physics distant from observations, and
that there is little reason to abandon hope that a solution exists.


\end{document}